\documentclass[conference]{IEEEtran}
\IEEEoverridecommandlockouts
\usepackage{cite}
\usepackage{adjustbox}
\usepackage{amsmath,amssymb,amsfonts}
\usepackage{graphicx}
\usepackage{textcomp}
\usepackage{xcolor}
\usepackage{multicol}
\usepackage{multirow}
\usepackage{subcaption}
\usepackage{enumitem}
\usepackage{balance}
\usepackage{booktabs}
\usepackage{array}

\usepackage{pst-all}
\usepackage{pstricks-add}
\usepackage{float}

\usepackage{tikz}
\usepackage{tikzscale}
\usepackage{transparent}

\usepackage{stackengine}

\newcommand\blfootnote[1]{%
  \begingroup
  \renewcommand\thefootnote{}\footnote{#1}%
  \addtocounter{footnote}{-1}%
  \endgroup
}

\usepackage{pgfplots}
\pgfplotsset{compat=1.13}
\usetikzlibrary{arrows.meta}

\usepackage{algorithm}
\usepackage[noend]{algpseudocode}

\usepackage{comment}

\definecolor{purplelight}{HTML}{ead1dc}
\definecolor{purpledark}{HTML}{a64d79}

\definecolor{purple}{HTML}{733d87}

\definecolor{colorRed}{HTML}{E19794}
\definecolor{colorGreen}{HTML}{b6d7a8}
\definecolor{colorBlue}{HTML}{a4c2f4}

\def\BibTeX{{\rm B\kern-.05em{\sc i\kern-.025em b}\kern-.08em
    T\kern-.1667em\lower.7ex\hbox{E}\kern-.125emX}}

\begin{document}

\title{A Trustworthy Recruitment Process for Spatial Mobile Crowdsourcing in Large-scale Social IoT \vspace{-0.0cm}}

\author{\IEEEauthorblockN{Abdullah Khanfor, Aymen Hamrouni, Hakim Ghazzai, Ye Yang, and Yehia Massoud}
\IEEEauthorblockA{
\small School of Systems  \& Enterprises, Stevens Institute of Technology, Hoboken, NJ, USA \\
Email: \{akhanfor, ahamroun, hghazzai, yyang, ymassoud\}@stevens.edu}
}
\maketitle
\thispagestyle{empty}

\begin{abstract}
\boldmath{
Spatial Mobile Crowdsourcing (SMCS) can be leveraged by exploiting the capabilities of the Social Internet-of-Things (SIoT) to execute spatial tasks. Typically, in SMCS, a task requester aims to recruit a subset of IoT devices and commission them to travel to the task location. However, because of the exponential increase of IoT networks and their diversified devices (e.g., multiple brands, different communication channels, etc.), recruiting the appropriate devices/workers is becoming a challenging task. To this end, in this paper, we develop a recruitment process for SMCS platforms using automated SIoT service discovery to select trustworthy workers satisfying the requester requirements. The method we purpose includes mainly two stages: 1) a worker filtering stage, aiming at reducing the workers' search space to a subset of potential trustworthy candidates using the Louvain community detection algorithm (CD) applied to SIoT relation graphs. Next, 2) a selection process stage that uses an Integer Linear Program (ILP) to determine the final set of selected devices/workers. The ILP maximizes a worker efficiency metric incorporating the skills/specs level, recruitment cost, and trustworthiness level of the recruited IoT devices. Selected experiments analyze the performance of the proposed CD-ILP algorithm using a real-world dataset and show its superiority in providing an effective recruitment strategy compared to an existing stochastic algorithm.
}
\end{abstract}

\blfootnote{\hrule
\vspace{0.2cm} This paper is accepted for publication in IEEE Technology \& Engineering Management Conference (TEMSCON '20), Detroit, MI, USA, Jun. 2020. \newline \textcopyright 2020 IEEE. Personal use of this material is permitted. Permission from IEEE must be obtained for all other uses, in any current or future media, including reprinting/republishing this material for advertising or promotional purposes, creating new collective works, for resale or redistribution to servers or lists, or reuse of any copyrighted component of this work in other works.}%


\vspace{-0.2cm}
\section{Introduction}
\vspace{-0.1cm}
The Internet-of-Things (IoT) is a broad network with heterogeneous connected smart entities (e.g., smartphones, autonomous vehicles, drones, etc.)~\cite{al2015internet,7491206}. These IoT entities are diverse with varied and diversified features (e.g., computational and storage capacities, communication protocols, etc.). They are embedded with numerous sensors such as GPS, magnetometer, and camera, allowing them to operate for a variety of applications~\cite{dweekat2018iot,khanfor2020automated,morienyane2019technology,khanfor2019application}. Spatial Mobile Crowdsourcing (SMCS) is an emerging IoT-based technology providing people the opportunity to take advantage of wearable and mobile phones' features. It refers to the technology that provides some IoT devices, called task requesters, to gain the benefit of the sensors of other IoT devices, called workers, such as wearable and mobile phones, and use them to execute specific sensing tasks. This collective power of the crowd can provide benefits for different levels of users in the IoT system, including the participants and the decision-makers~\cite{jian2015mobile}. One of the most simplistic examples of IoT-based SMCS systems is to manage and handle sensor-collected data. A user (e.g., local authority) could tap into an application and submit the locations to be monitored~\cite{8982179}. Mobile IoT workers can spatially travel to the requested locations and collect the required data and forward it to the requester.

In general, in typical SMCS framework~\cite{article5}, two external agents are interacting with the cloud platform: the task requesters and the workers. The task requester, which can be a human carrying a smartphone as well as other IoT devices (e.g., autonomous vehicle), provides its task information and requirements to the platform so it can be announced and executed by the crowd of IoT devices, e.g., collecting photos or sensing data. The cloud platform hosting the main framework then uses these criteria to recruit suitable workers capable of traveling to the tasks' locations to execute the tasks and upload their results to the platform. The latter pre-processes the data and sends the final results to the task requester. Because of the diversity of IoT devices and workers, they can provide different response quality. Moreover, it can be incentivized differently to compensate for their achievements. Thus, the selection of IoT workers is not a straightforward process. Most of the recent SMCS studies, for example, ~\cite{7446292,8100997,9028164}, focus on optimizing the recruitment process by hiring skilled workers for each task such that they can fulfill the tasks' requirements and provide suitable results.

The level of trustworthiness between the requester and workers is one of the vital criteria to achieve satisfactory outcomes and ensure a certain privacy and security levels. Usually, the requester prefers trustworthy and reliable workers to perform the tasks. Therefore, in this paper, we propose to take advantage of the social IoT (SIoT) concept, where IoT objects can establish social relations~\cite{atzori2012social}, to leverage SMCS recruitment. The social relationships between IoT objects can be built according to the communication links, owners' policies, and interactions between objects. 
SIoT relations can be effectively exploited in the SMCS context to ensure trustworthiness when executing crowdsourcing tasks~\cite{wang2016toward}.

In this paper, we propose to develop an effective SMCS recruitment process in large-scale IoT networks by selecting appropriate, skilled, low-cost, and trustworthy IoT workers. We aim to adopt a two-phases approach. First, we proceed with a worker pre-processing phase where the objective is to exploit the SIoT relationships among IoT devices and task requesters to select trustworthy pre-selected candidates. Two types of social relationships are considered in this case: an owner's social network-based relation and automated built social object relation. A community detection (CD) algorithm applied to the graphs representing the social relationships among devices in the area of interest to determine a subset of trustworthy potential candidates that can be recruited. Then, an integer linear programming (ILP) is then formulated and solved to hire the skilled and socially connected hired workers to complete the SMCS tasks.

The CD-ILP will hire skilled and socially connected workers to complete SMCS tasks. In summary, the proposed CD-ILP is composed of two components, and its work-flow is as follows: i) the community detection component uses the Louvain community detection algorithm \cite{blondel2008fast} to reduce the search space of devices in the SIoT graph. ii) Then, an ILP is applied to considers only candidates using three main key selection metrics: 1) required skills, 2) budget allocation, and 3) the task requester relationships with devices (workers). As a bench-marking algorithm, we implement a heuristic stochastic approach already discussed in the literature based on the optimal stopping strategies \cite{793723} and compare its performances with our proposed CD-ILP approach. The results of the conducted experiments between the stochastic algorithm applied to the initial dataset and the CD-ILP algorithm show that our proposed approach achieves better recruitment performances with lower computational complexity.

\vspace{-0.0cm}
\section{SMCS Framework}
\begin{figure}[!t]
\vspace{0.3cm}
    \centering
    \hspace{-0.5cm}
    \centering
    \includegraphics[width=\columnwidth]{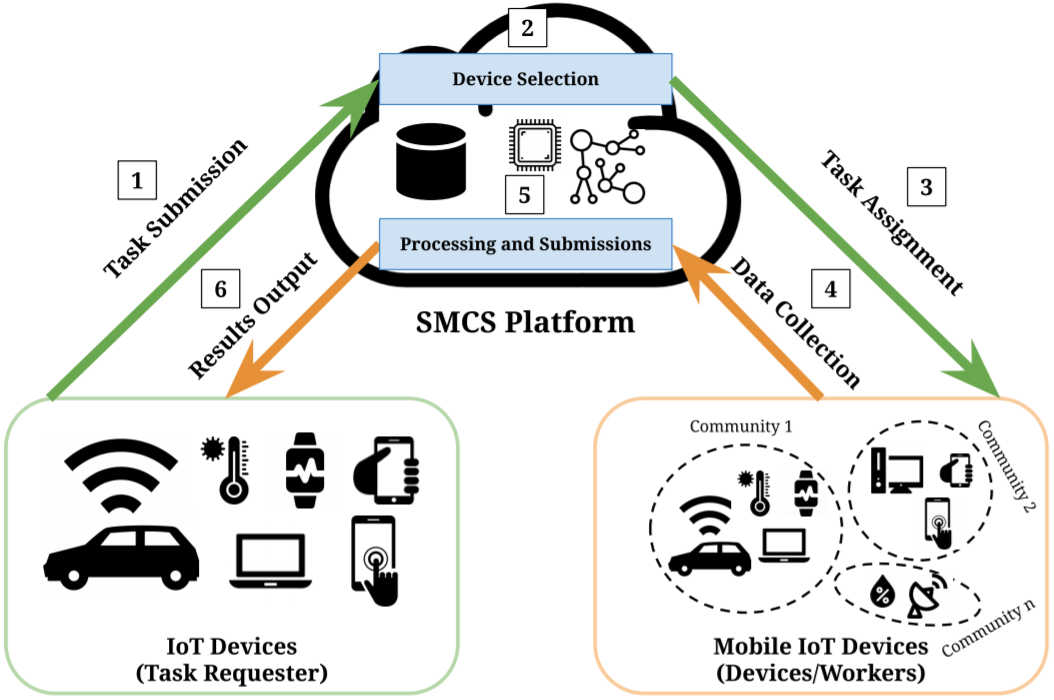}
    \caption{SMCS High-level Architecture.}\vspace{-0.15cm}
    \label{fig:arch}
\end{figure}

\begin{figure*}[!h]\vspace{0.3cm}
    \centering
    \includegraphics[width=0.95\textwidth]{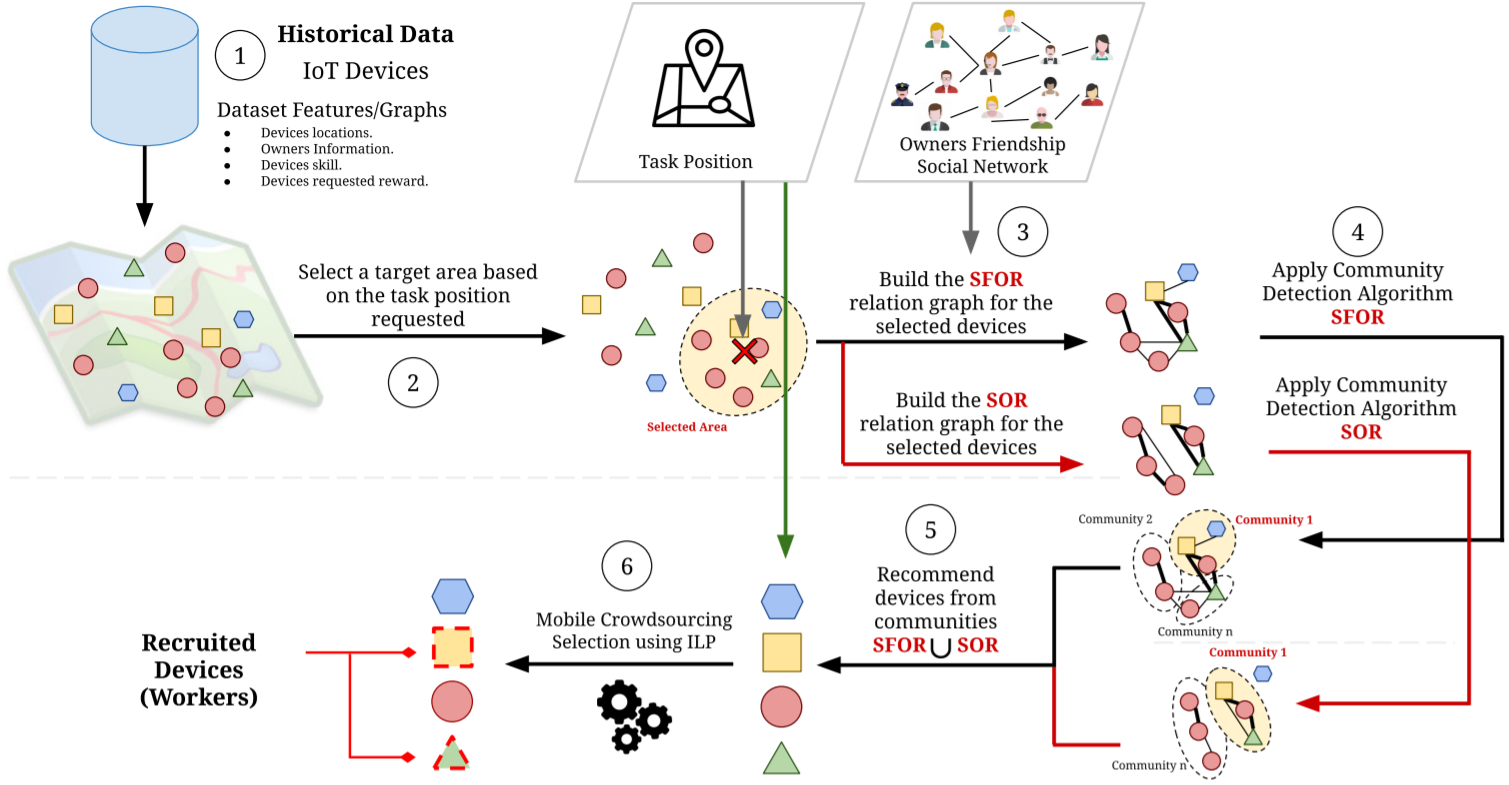}\vspace{-0.1cm}
    \caption{Proposed Recruitment Process for large-scale SIoT network.}
    \label{fig:framework}\vspace{-0.2cm}
\end{figure*}

\vspace{-0.0cm}
The main components of the SMCS framework that interact with the cloud platform are illustrated in Fig.~\ref{fig:arch}. These components are the task requester (e.g., IoT device, human, etc.) and the available workers (e.g., smartphones carried by humans, autonomous vehicles, sensors, etc.). When it needs services, the task requester submits its SMCS task $t$ with its owner preference $own_t$ for the workers to the platform. Also, it defines the location $loc_t$ of the task along with the set $\mathcal S_t$ of required skills/specs (e.g., expertise for humans, device specifications, etc.). The centralized architecture of the framework considers all the available workers as candidates and search for a suitable set of devices capable of satisfying delivery results. This can be time-consuming and causes a useless overflow to the server. To this end, we propose our framework, illustrated in Fig.~\ref{fig:framework}. The proposed method comprises a filtering process phase that is controlled by the SMCS platform. The initially available workers $N$ goes through a selection mechanism for spatial filtering. The latter selects a suitable subset of workers within a radius $R$ of the task position $loc_t$ that the requester pre-defined. Afterward, the devices' ownership and its owner's social network are used to establish a graph with several relations between devices. At the same time, the devices that are socially exposed to each other will have links (i.e., relationships). Next, a CD algorithm is applied to maintain a subset of trustworthy potential workers from which the platform will hire and select the final workers by optimally an ILP. In the following sections, we will explain the spatial filtering procedure, the social graph creation, the CD technique as well as the final worker selection process by going through each stage of the framework denoted by steps 1 to 6 in Fig.~\ref{fig:framework}.

\section{Worker Filtering Process}\label{step1}
The filtering aims to prepare and select the devices/workers for the next stage of the SMCS recruitment process. The filtering includes the objects selection, discussed in the following Section~\ref{sec:obj_selection}, based on the task position $loc_t$. Next, building the relations between the devices in Section~\ref{sec:relation} to apply the CD algorithm to get the clusters of trustworthy devices.
\subsection{Objects Selection}\label{sec:obj_selection}
The task location is designated by the requester $req_i$ of a task. Also, the IoT devices include their geographical positions on the map. The device selection process starts by measuring the distance between the task position $loc_t$  and the devices' locations $N$. The devices within a range of less than $R$ from the requester's location are kept for the pre-processing phase. This process will require $O( N )$. The location-based filtering is illustrated in steps 1 and 2 in Fig. \ref{fig:framework}. 
\subsection{Objects Relations}\label{sec:relation}
This system includes various devices such as smartphones, smartwatches, weather sensors, and personal computer devices, etc. To establish the relationships and maintain a certain level of trustworthiness between the devices, we consider the following social relations in SIoT context:

$\bullet$ \textit{Social friendship and ownership relation (SFOR):} This relationship is formed with consideration of the IoT device ownerships and the owners ' social friendships and networks. First, a set of devices will produce a fully connected subgraph if they are owned by one entity. It is used to determine IoT devices having relations reflecting the relationships between their owners. Next, the owner's social relationships, such as friendship and co-operators, can allow authorized entities to gain access rights and therefore, can be considered as socially connected IoT objects. Not only that, mutual friends, \textit{aka} friend of a friend, in a social network, can grant some access to each other and, therefore, degraded strength of links established between friends based on the number of nodes between each friend in the social network.

$\bullet$ \textit{Social object relationship (SOR):} If two IoT entities come into contact, the link is formed. This relation could be either infrequent or continuous, depending on device owner policy requirements.

In Fig. \ref{fig:framework}, stated in step 3, the relations are established between devices based on the previous definitions. Then, relations graphs are created for each social relationship where a CD algorithm is applied to distinguish multiple communities where IoT devices share strong relationships with each other or with the task requester.
\subsection{Community Detection Algorithm}\label{step2}
In the fourth step of the proposed recruitment framework, we apply a community detection algorithm using the already defined relations. In network analysis, communities are identified if there is a collection of nodes that are connected strongly compared to the rest of the links of the network~\cite{flake2004graph}. One of the primary benefits of identifying communities can be used to improve the information retrieval of the network. Likewise, it helps to reduce the search space instead of scanning all graph nodes, in our context the devices/workers, by narrowing it to the desired community and produce a small set of workers $\mathcal W_t$. Thus, it offers a process for automated service discovery in large-scale IoT systems.

We use the Louvain algorithm for the simplicity of implementation and the low running time complexity in a vast scale network such as SIoT. Louvain is one of the fastest algorithms to identify non-overlapping, \textit{aka} disjoint, communities. The technique is a greedy method with a running time of $O(n \log n)$. It focuses on the modularity score to identify the different communities. The modularity score is to show the value of node allocation to the community by analyzing the edge density within a collection of nodes relative to how it would be linked to a random network. Mainly, it is trying to maximize the modularity score for each community. At the end of this phase, the potential set of workers $W_t$ are the result of the community detection component representing the union of the SFOR and SOR and having a certain level of trustworthiness defined by the task requester (i.e., they have a strong connection with the task requester).

\section{Selection Process} \label{step3}
After completing the filtering phase, the initial number of workers $N$ is reduced to a smaller set given the location $loc_t$ of task $t$ provided $req$. 
Let $\mathcal W_t=\{1, \dots, W_t\}$ be that set of candidate workers for task $t$. Given the set $\mathcal S=\{1,\dots, S\}$ of all $S$ possible skills in the system, we define the logical skill quantity for a task $t$ by $Q_t(s),\,s\ \in \mathcal S$ where $Q_t(s)=1$ if the skill $s$ is required by task $t$ and $Q_t(s)=0$ otherwise. Hence, the skills set required by the task $t$ is $\mathcal S_t=\{s \in \mathcal S/Q_t(s)=1\}$. Each worker $w \in \mathcal W_t$ has a degree of expertise in skill $s \in \mathcal S$ denoted by $S_{ws}$ where $0\leq S_{ws} \leq 1$. The term $S_{ws}$ represents the expertise value of skill $s$ that worker $w$ has and it is interpreted as follows: $S_{ws} \leftarrow 1$ means that the worker $w$ is an expert in skill $s$. Otherwise, $S_{ws}\rightarrow 0$. We suppose that each recruited worker can only contribute with one required skill. Consequently, for a task $t$ having as a skill set $\mathcal S_t$, the number of recruited workers must be $|\mathcal S_t|$.  To execute a task with skill $s$, a worker $w$ may request a certain cost denoted by ~$C_{ws}$.

Consequently, the efficiency of worker $w$ chosen to contribute in task $t$ with skill $s$ is written as follows: 
\begin{align}
E^t_{w,s}= & \eta_1\frac{ S^t_{w,s}}{\bar{S}}  -   \eta_2 \frac{C^t_{w,s}}{\bar{ C }}  +   \eta_3 \frac{O_{t,w}}{\bar{O}}.
\label{ppp}
\end{align}
The efficiency expression which the SMCS aims to maximize is established using three key metrics. The first metric is $S_{w,s}$, and it represents the skill level in $s$ of IoT device $w$. The term $C_{w,s}$ represents the cost of worker $w$ providing skills $s$, and it can be expressed by: 
\begin{align}
C^t_{w,s}= R_{w,s} + \Delta_{t,w} \times P,
\label{ration}
\end{align}
where 
$R_{w,s}$ represents the cost demanded by worker $w$ when providing the skill $s$. The term $P$ is a coefficient defined by the platform that converts the unit of traveled distance to a reward unit, e.g., monetary unit (MU). 
The variable $\Delta_{t,w}$ represents the distance separating the two locations $loc_w$ and $loc_t$. The last term $O_{t,w}$ in \eqref{ppp} represents the ownership distance between the task requester and worker, and it is introduced to ensure a certain trustworthy level between the recruited devices and the task requester. The quantities $\bar{X}$ in the denominator of each term of \eqref{ppp} are introduced for normalization purposes so that the four key metrics have the same order of magnitude. The weights $\eta_f$, with $f \in \{1,2,3\}$ and $\sum_{f=1}^{3} \eta_f=1$, indicate the SMCS platform recruitment strategy. For example, the case when the platform's strategy is only to recruit skilled workers (i.e., $\eta_1=1$ and $\eta_2=0$, $\eta_3=0$.)

To indicate the worker $w$ assigned for a task $t$ to contributed with skill $s$, we introduce a binary decision variable $x^t_{ws}$ defined as follows:
\begin{align} \label{x}
x^t_{ws}= &
     \begin{cases}
       \text{1,} &\,\text{if worker $w$ is chosen to contribute in task $t$}\\
       &\text{with skill $s$,}\\
       \text{0,} &\,\text{otherwise,} \\ 
     \end{cases}\notag\\
 &\hspace{2cm}\forall \, w \mathcal \in W_t, \, \forall \, s \in \mathcal S_t,\text{ and }  \forall \, t \in \mathcal T.
     \end{align}
Also, we present all the constraints required for an optimal selection respecting both task requesters' and workers' demands.

$\bullet$ \textbf{Mono Task-Skill Constraints:} These constraints are added to ensure that a worker $w$ can be chosen to complete at most a task $s$ and can only contribute with at most a skill $s$.
These constraints are presented as follows:
\begin{align}
\sum_{w\in \mathcal W_t} \sum_{s \in \mathcal S_t} x^t_{ws}  \leq 1, \forall \, t \in \mathcal T,  \label{onetask} \\
\sum_{t \in \mathcal T} \sum_{w \in \mathcal W_t} x^t_{ws}  \leq 1, \forall \, s \in \mathcal S_t.\label{oneskill}
\end{align}

Constraint \eqref{onetask} forces the optimizer to recruit worker $w$ to contribute in at most one task $t$.
Constraint \eqref{oneskill} forces each worker $w$ to provide at most one skill $s$ for the task $t$.

$\bullet$ \textbf{Tasks' Skills Fulfillment Constraint:}
The following constraint ensures that each of the hired workers for task $t$ contributes with a required skill defined in the task:
\begin{align}
&\hspace{-0.4cm}\sum_{t \in \mathcal T } \sum_{w  \in \mathcal W}^{}x^t_{ws} = Q_t(s), \, \forall \, s \in \mathcal S_t, \forall \, t \in \mathcal T,\label{requiredskill}
\end{align}

The selection optimization problem in SMCS is then defined as follows:
\begin{align}
\text{(P):} & \underset{ {x^t_{ws} \in \{0,1\}}}{\text{ maximize }}
\sum_{w \in \mathcal W_t} \sum_{s \in \mathcal S_t}^{}  \sum_{t \in \mathcal T}^{} & x^t_{ws} \bigg [ \eta_1\frac{ S^t_{w,s}}{\bar{S}}  -   \eta_2 \frac{C^t_{w,s}}{\bar{ C }}  +  \notag \\ &  &\eta_3 \frac{O_{t,w}}{\bar{O}} \bigg ] ,  \notag \\ 
&\text{subject to:} \notag \\
&\eqref{onetask},\,\eqref{oneskill},\text{ and } \eqref{requiredskill}. \notag
\end{align}

This optimization problem in (P) is formulated as an ILP, and the solution can be optimally obtained using off-the-shelf software integrating the branch and bound algorithms and simplex method. The output of (P)  determines the exact workers hired by the platform to execute the required task with maximum worker efficiency level. Note that there are cases where the problem is infeasible, for example, when the set of worker $\mathcal W_t$ obtained from the pre-processing phase is empty or the number of workers $W_t$ is relatively less than the number of skills $|\mathcal S_t|$. However, in real life SMCS platforms, this case is unlikely to occur since, by definition, in large-scale IoT systems, the value of $W_t \gg  |\mathcal S_t|$, $\forall t \in \mathcal T$. 
If not, the platform can increase the search radius $R$ and repeat the SMCS solving process.

\section{Simulation and Evaluation}\label{sec4}

In this section, we study the behavior of the proposed community detection strategy and the ILP-based selection model. We evaluate the complete work-flow performances using various metrics and compare them with the ones of a stochastic approach, which is based on the odds-algorithm and uses the optimal stopping strategies \cite{793723} to compute its output. It consists of making a decision by observing multiple workers set one after the other and stopping on the first interesting worker set. Moreover, the optimal stopping rule prescribes always rejecting the first $30\%$ workers set that are tested and then stopping at the first workers set which is better at (P) than every workers set interviewed so far (or continuing to the last possible workers set if this never occurs).

\subsection{Dataset and Experimental Setup}
\begin{figure}[!h]
\vspace{0.2cm}
    \centering
    \includegraphics[width=\columnwidth]{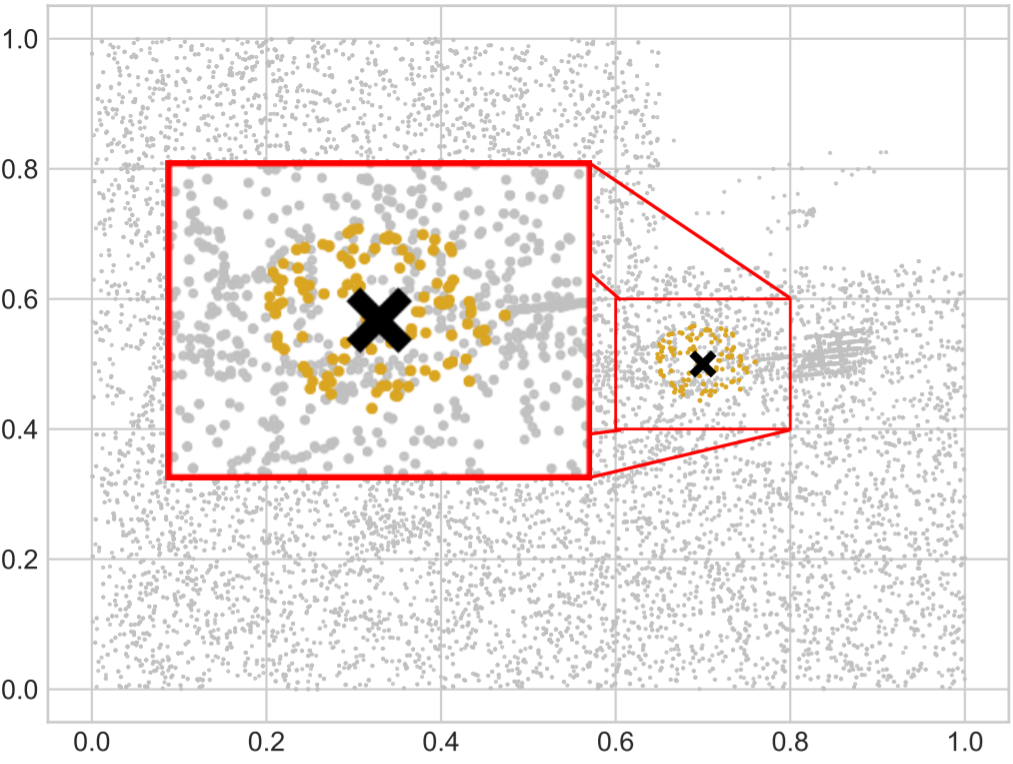}\vspace{-0.0cm}
    \caption{Normalized locations of IoT devices. The black 'X' mark and the red box represent the task location and the investigated area, respectively.}
    \label{fig:selectedArea}\vspace{-0.3cm}
\end{figure}

The dataset~\cite{marche2018dataset} used in this experiment includes real-world IoT artifacts obtained and modeled in Santander, Spain. The total number of objects is 16216, comprising 14600 private users and 1616 public service IoT devices. We select a small area to demonstrate our framework. We need to reduce the number of devices significantly and ensure a fast running time of the framework. Within the small space, we calculate the distance between the requester position and all the devices in the area. If the distance range is above a certain threshold, then the device is ignored. Fig.~\ref{fig:selectedArea} displays the selected area in the red rectangle, the requester position with x mark in black, and the potential candidates within the specified radius from the requester are denoted with yellow dots to indicate the possible devices that will be recruited.

For the experimental setup, we set $S=5$ and $T=20$. We perform Monte Carlo simulations where $1000$ iterations of different parameter settings are generated, and results are averaged upon them. We also adopt a proportional selection strategy and set $\eta=\frac{1}{3}$, $\forall i \in \{1,2,3\}$. In our experiments, all algorithms are implemented in a Python 3.6 environment and run on a 32 socket Intel(R) Xeon (R) E5-2698 v3 @2.30GHz CPU with 48GB of RAM. To solve the ILP algorithm, we use the Python API of academical CPLEX.
\vspace{-0.2cm}
\subsection{Relations Formation and Community Detection}
Since we lack the owner's social network, we create a social network between the owners using the Watts-Strogatz model~\cite{watts1998collective} a random graph generation. In this model, we set the parameters for the number of nodes that reflects the number of owners in the dataset, excluding the public-owned devices.  Besides that, we set $p$ the probability of adding a new edge to $p=0.5$ to represents a small-world network between the friends between the owners. The social network between the owners helps us form the definition of SFOR in Section~\ref{sec:relation}.

\begin{figure}[!h]
    \centering
    \includegraphics[width=0.9\columnwidth]{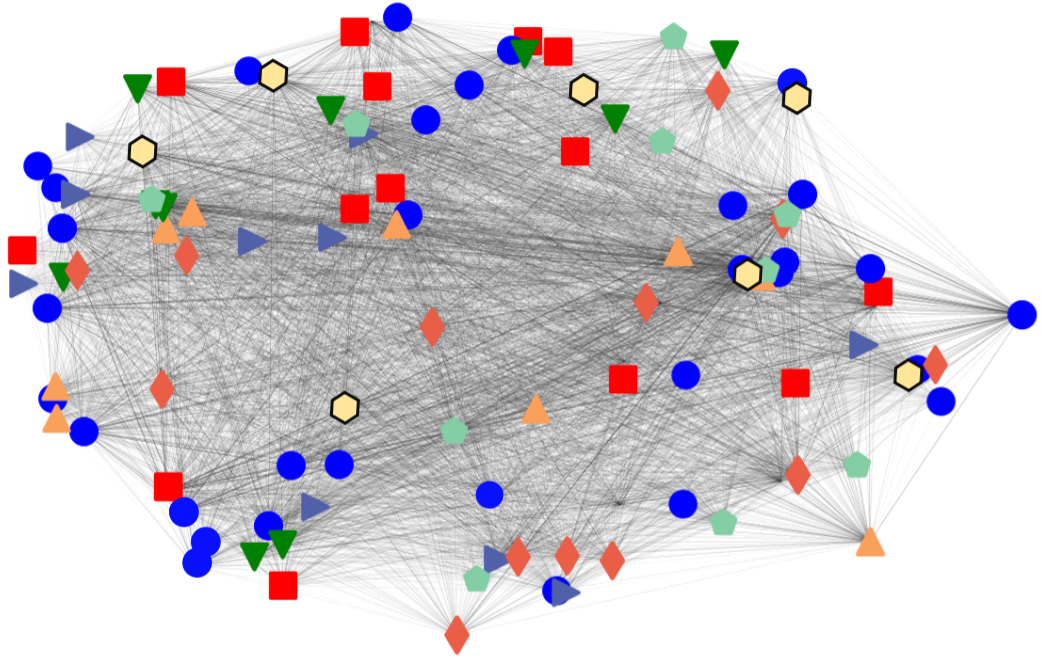}\vspace{-0.0cm}
    \caption{Visualization of 11 detected communities using SFOR relation. Nodes having same color and shape belong to the same community.}
    \label{fig:SFORViz}\vspace{-0.3cm}
\end{figure}

We apply the Louvain algorithm on the SFOR relation on the reduced space produced from the previous process, and we visualize the results in Fig. \ref{fig:SFORViz}. We notice that there are 11 communities in the small network, and they are labeled by different colors and shapes using displayed in Fig. \ref{fig:SFORViz}.

\begin{figure}[t]
\vspace{0.2cm}
    \centering
    \includegraphics[width=0.8\columnwidth]{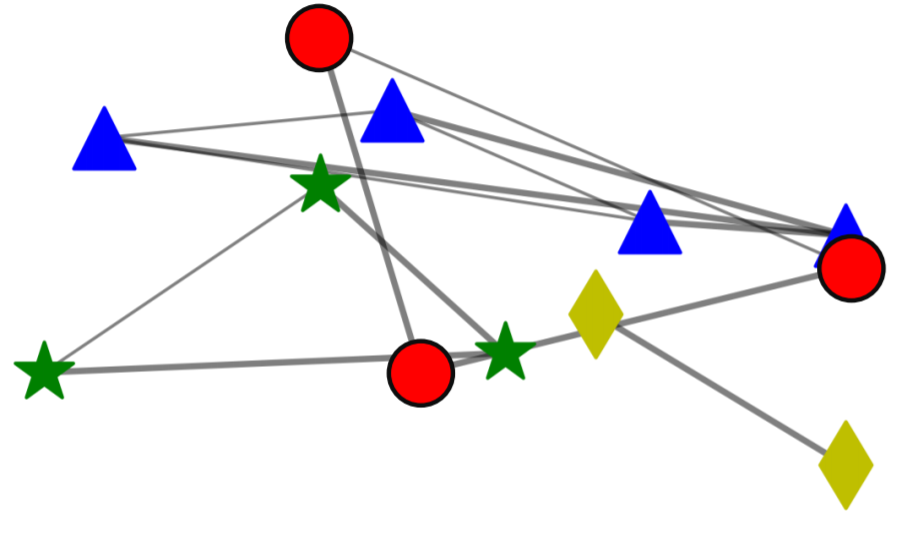}\vspace{-0.0cm}
    \caption{Visualization of 4 detected communities using SOR relation. Nodes having same color and shape belong to the same community.}
    \label{fig:SORViz}
    \vspace{-0.4cm}
\end{figure}

For the SOR, that is provided by Marche et al. \cite{marche2018dataset} dataset. It is built based on three parameters: the number of meetings is three or more, the period of the meeting is 30 minutes, and the time between two successive meetings is at least 6 hours. We visualize the results of the small selected area in Fig. \ref{fig:SORViz}. We observe a few devices that are fully connected, which form cliques.

\vspace{-0.2cm}
\subsection{Selection Simulation}
\begin{figure*}[t]
    \centering
    \includegraphics[width=\textwidth]{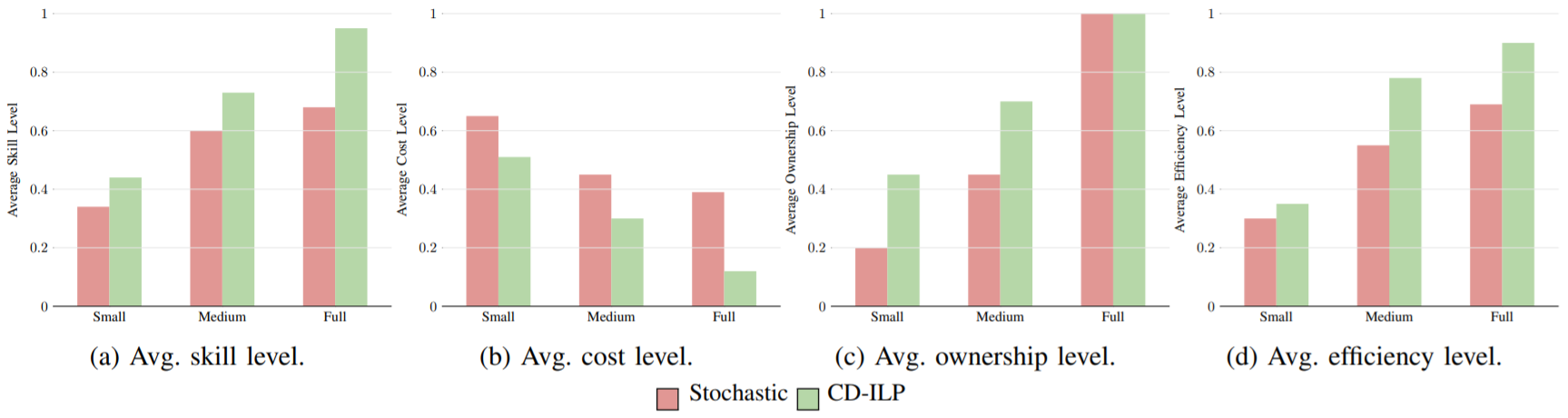}\vspace{-0.0cm}
    \caption{Average skill, cost, ownership, and efficiency (objective function) of the recruitment simulation with three different social network connectivity scales (small, medium, and full). }
    \label{fig:SCOEConnctivity}
\end{figure*}
\begin{figure}[!h]
    \centering
  \includegraphics[width=\columnwidth]{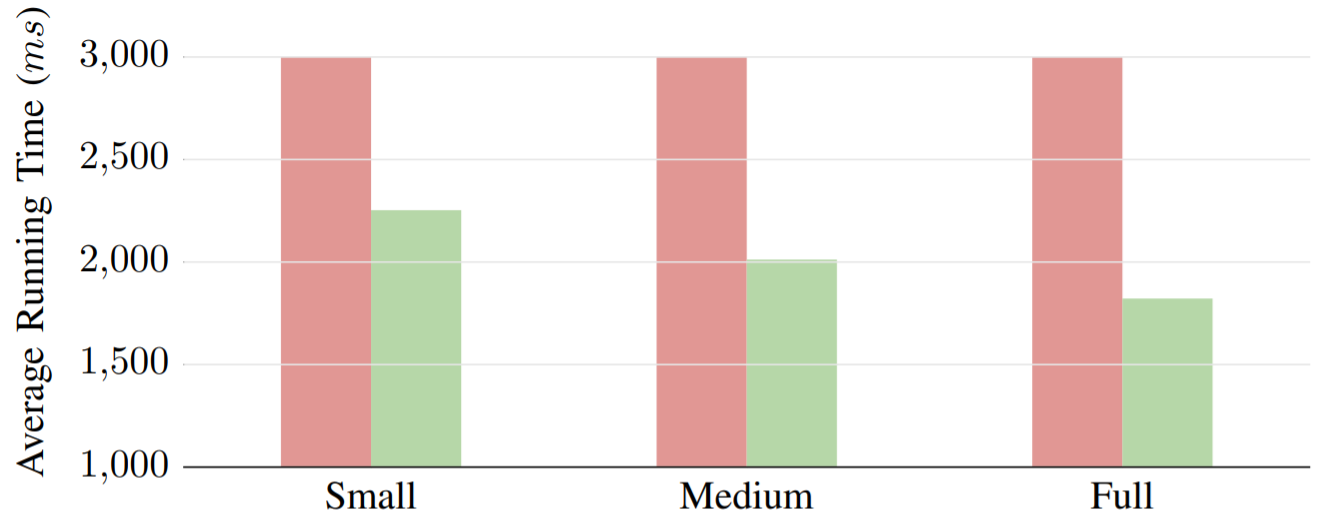}\vspace{-0.0cm}
  \caption{Average running time ($ms$) vs. ownership degree for the stochastic algorithm (red bars) and the CD-ILP (green bars).}
  \label{fig11}
  \vspace{-0.0cm}
\end{figure}

To evaluate the performance of the proposed CD-ILP approach, we conduct two main simulations. For the first one, we compare the performances of our proposed CD-ILP algorithm against the implemented stochastic approach. 
As shown in Fig.~6 (a) through Fig.~6 (d), we perform an average evaluation of the selected workers while varying the size of IoT devices' owner social network for each task using the following five metrics: overall efficiency, skills efficiency, workers' cost, ownership level, and running time. It is shown that, for each IoT device's owner's social network size, the CD-ILP approach achieves better performances than the existing stochastic model. In fact, for example, the cost of the selected workers and their skill level using the CD-ILP approach is slightly higher than the stochastic one. Furthermore, for the skill level simulation, as illustrated in Fig.~6 (a), we notice that the performances increase while expanding the size of the IoT device's owner's social network. This is explained by the fact that both algorithms achieve better solutions when the device's owner SIoT network expands to include other IoT devices (i.e., the owner's network includes even the friend-of-a-friend relations). The same behavior and interpretation can be noticed for the workers' cost shown in Fig.~6 (b) and for the overall efficiency illustrated in Fig.~6 (c).

The running time for both algorithms, as shown in Fig.~\ref{fig11}, indicates that the CD-ILP algorithm has a lower running time compared to the constant behavior of the heuristic stochastic bench-marking algorithm. We also notice that, for these metrics, the gap between the algorithms increases while increasing the size of the IoT devices' owner social network. This can be explained by the fact that, when expanding the size of IoT devices owner's social network, the performances of the CD-ILP improves. Further, the stochastic complexity time's performances remain constant because the number of iterations achieved by the heuristic approach is invariant to the network size.

\begin{figure}[t]
  \centering
  \includegraphics[width=\columnwidth]{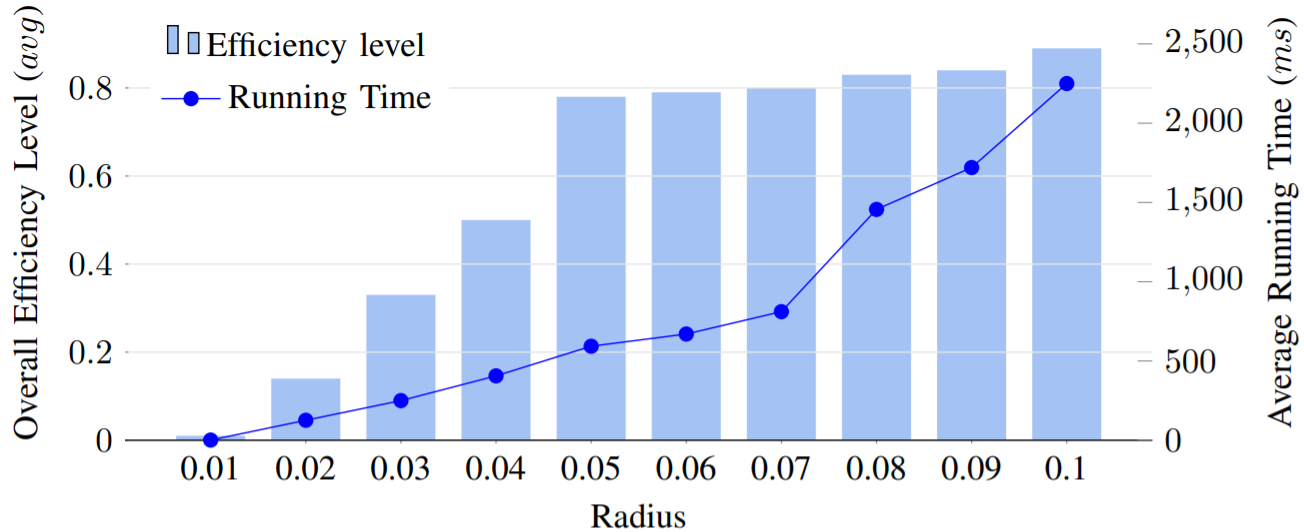}\vspace{-0.0cm}
  \caption{Overall efficiency level (bars) and average running time (solid line) for the CD-ILP algorithm vs. the circle radius $R$.}
  \vspace{-0.3cm}
  \label{fig10}
\end{figure}

The second simulation is conducted using the proposed CD-ILP algorithm. It brings out the effect of varying the radius of the circle $R$ as illustrated in Fig.~\ref{fig10}; the overall efficiency of the selected workers using the CD-ILP algorithm increases with square root growth, unlike the average running time, which increases exponentially. Consequently, we can suggest that increasing the radius beyond $50\%$ increases the average running time without significant improvement of the overall efficiency.

\vspace{-0.1cm}
\section{Conclusion}
In this paper, we developed a formulation for the spatial recruitment process in SMCS platforms using SIoT systems. The proposed approach involves a community detection technique that reduces the initial set of workers to a set of recommended devices suitable and trustworthy to execute the task. Then, an ILP is run on the candidate set to optimize and output the most appropriate set of workers. The results of the conducted experiments show that the CD-ILP algorithm outperforms the bench-marking stochastic algorithm, applied to the initial dataset, with lower computational complexity.

\balance
\bibliographystyle{IEEEtran}
\bibliography{references}

\end{document}